\documentclass[12pt]{article}
\usepackage{epsfig}

\begin{document}

\titlepage
\begin{center}
{\bf \Large Higher order corrections to primordial spectra from 
cosmological inflation}

\end{center}

\vskip 1cm

\begin{center}
{\large Dominik J. Schwarz $^a$
\footnote{dschwarz@hep.itp.tuwien.ac.at},
        C\'esar A. Terrero-Escalante $^b$
\footnote{cterrero@fis.cinvestav.mx}
        and Alberto A. Garc\'{\i}a $^b$
\footnote{aagarcia@fis.cinvestav.mx}
}
\end{center}
\vskip 0.5cm
\begin{center}
$^a$ {\small 
\it Institut f\"ur Theoretische Physik, Technische Universit\"at Wien, \\
Wiedner Haupstra\ss e 8--10, 1040 Wien, Austria.}\\
$^b$ {\small 
\it Departamento de F\'{\i}sica,
Centro de Investigaci\'on y de Estudios Avanzados del IPN,
Apdo.~Postal 14-740, 07000, M\'exico D.F., M\'exico.}
\end{center}

\vskip 2cm
\begin{center}
{\large Abstract}
\end{center}
\noindent
We calculate power spectra of cosmological perturbations at high accuracy 
for two classes of inflation models. We classify the models according to the
behaviour of the Hubble distance during inflation. Our approximation works 
if the Hubble distance can be approximated either to be a constant or to 
grow linearly with cosmic time. Many popular inflationary models can be 
described in this way, e.g., chaotic inflation with a monomial potential,
power-law inflation and inflation at a maximum. Our scheme of approximation 
does not rely on a slow-roll expansion. Thus we can make accurate predictions 
for some of the models with large slow-roll parameters. 


\newpage

\section{Introduction}

The most important prediction of cosmological inflation \cite{inflation} 
(besides the spatial flatness of the universe) is the generation of 
primordial fluctuations of matter and space-time \cite{perturbations}. 
These fluctuations seed the formation of large scale structure and 
give rise to anisotropies in the cosmic microwave background (CMB). 
Recent CMB observations \cite{CMBdata} have clearly detected the first 
three acoustic peaks. In order to explain these observations a 
dominant contribution of adiabatic perturbations is necessary,
as predicted by cosmological inflation. The position of these 
peaks is consistent with a spatially flat universe. 
The level of accuracy of current experiments is of the order of $10\%$,
which roughly corresponds to $10\%$ uncertainty for the best determined 
cosmological parameters. Future experiments \cite{future} will increase 
this accuracy to the limit of the cosmic variance (of the order of $1\%$ 
at the arcminute scale). 

For a reliable comparison of inflationary predictions and CMB data we need  
analytical predictions that can meet the accuracy of the observations. 
The primordial fluctuations are characterised in terms of 
power spectra of scalar and tensor perturbations (inflation driven by 
a scalar field predicts that there are no vector perturbations). 
Only a limited number of very special inflationary models is known  
for which the power spectra can be calculated exactly \cite{MS3}. Thus we
have to use some approximation or we have to rely on numerical 
calculations \cite{GL2}. 
The state of the art of the analytical calculation of the spectral 
amplitudes are the approximate expressions due to Stewart and Lyth \cite{SL}, 
which are first order expressions in terms of the so-called, slow-roll 
parameters. These 
expressions have been tested in Refs.~\cite{GL,MS2} to be precise enough 
to match the accuracy of current observations, given the conditions for 
slow roll are met. However, it has been pointed out that for the 
analysis of data from future observations, more precise analytical 
expressions are compulsory \cite{MS2}. Moreover, there are inflationary 
models that do not belong to the class of slow-roll models.
The approach by Stewart and Lyth \cite{SL} includes several 
approximations.
In a first step the mode equations of the fluctuations are approximated by 
Bessel differential equations. With this aim, one has to assume that 
the slow-roll parameters are constant, which is a consistent procedure up to 
first order in the slow-roll parameters only. A second approximation is 
made by assuming that the 
dominant modes take their superhorizon values at the moment of horizon 
crossing, 
which is arbitrarily fixed as $k = aH$. Martin and Schwarz \cite{MS2} closed a 
gap in the derivation of this approximation, showing that the time of matching 
the Bessel function solution to the superhorizon solution has to coincide with 
the time when the Hubble rate and the slow-roll parameters are evaluated.
Wang, Mukhanov and Steinhardt \cite{WMS} have shown that the method of Stewart 
and Lyth \cite{SL} does not allow to improve the Bessel function approximation 
to higher orders.
Recently, Stewart and Gong \cite{GS} introduced a method to calculate the 
spectra to higher orders in the slow-roll parameters for slow-roll inflation. 
This new method does no longer refer to the Bessel function approximation, 
but instead uses an approximation based on Greens functions. 
They calculate 
the scalar spectral amplitude at second order in the slow-roll parameters. 

In this paper we present an approximation that does not rely on the 
slow-roll approximation and has the advantage that models for which some 
of the slow-roll parameters take large values can be treated. We calculate 
the amplitudes with high precision for two families of inflationary models, 
models that are characterised
by an almost constant Hubble distance during inflation and models that are 
characterised by an almost linearly growing Hubble distance. The common 
link between 
both families is that their spectral indices are almost constant, even for 
large 
values of some of the slow-roll parameters. 
Our approximation makes use of the 
Bessel function approximation, but avoids the argument by Wang, Mukhanov and 
Steinhardt \cite{WMS} by putting constraints on the relative magnitude
of the various slow-roll parameters.

The traditional set of slow-roll parameters (see e.g. \cite{Lea}) is not well
suited for the new approximation. Although we hesitated to introduce yet 
another set of parameters that control the dynamics of inflation, our new 
parameters have two major advantages. Firstly, their definition is simpler 
than other definitions of slow-roll parameters, and, secondly, the new 
definition 
allows a very transparent physical interpretation of the parameters:
the parameters control how the Hubble distance behaves during inflation and 
thus we call them the horizon flow functions (or parameters when we evaluate 
these functions at a certain moment of time). 
We have presented a variant of our new approach using the traditional notation 
in Ref.~\cite{TESGProc}. 

This paper is organised as follows: In section 2 we introduce the horizon flow 
functions and establish the link to the slow-roll parameters. The core of 
the paper is section 3, where we calculate the scalar power spectrum in the 
constant-horizon and in the growing-horizon approximations up to third order. 
The tensor power spectra are calculated in section 4. In section 5 we 
present the corresponding `consistency relations of inflation'.
We finally compare our 
results to the result of Stewart and Gong \cite{GS} and show that both 
results are consistent, which is a nontrivial check of both calculations.

\section{Horizon flow}

For the reasons explained in the introduction we define a set of horizon 
flow functions starting from
\begin{equation}
\epsilon_0 \equiv \frac{{d_{\rm H}}(N)}{d_{\rm Hi}},
\end{equation} 
where $d_{\rm H} \equiv 1/H$ denotes the Hubble distance, 
$N \equiv \ln (a/a_{\rm i})$ the number of e-folds since some initial time 
$t_{\rm i}$, and $d_{\rm Hi} \equiv d_{\rm H}(t_{\rm i})$. (Note that usually 
the 
number of e-folds is counted backward in time, we count it forward,
i.e., $N(t_{\rm i})=0$.) The quantity $d_{\rm H}$ is commonly called 
``horizon''
because it is a good estimate of the size of the region that may be in causal
contact within one expansion time.
 
Further we define a hierarchy of functions in a systematic way by
\begin{equation}
\label{eq:Hjf}
\epsilon_{m+1} \equiv {{\rm d} \ln |\epsilon_m|
\over {\rm d} N}, \qquad m\geq 0.
\end{equation} 
According to this definition,
\begin{equation}
\epsilon_1 \equiv {{\rm d} \ln d_{\rm H}\over {\rm d} N},
\end{equation} 
measures the logarithmic change of the Hubble distance per e-fold of 
expansion. 
Inflation happens for $\epsilon_1 < 1$ 
(equivalent to $\ddot{a} > 0$) and $\epsilon_1 >0$ from the weak energy 
condition (for a spatially flat universe). For $m>1$, $\epsilon_m$ may take 
any real value. Expressions (\ref{eq:Hjf}) define a flow in the space 
$\{ \epsilon_m\}$ with the cosmic time being the evolution parameter. This 
flow is described by the equations of motion
\begin{equation}
\label{eq:eom}
\epsilon_0\dot{\epsilon_m}-\frac 1{d_{\rm Hi}}\epsilon_m \epsilon_{m+1} = 0.
\end{equation}
For $m = 0(1)$ we find $\epsilon_1= \dot{d_{\rm H}}$ and 
$\epsilon_1\epsilon_2 =
d_{\rm H} \ddot{d_{\rm H}}$ respectively, which describe the time 
evolution of the horizon.

Let us stress the advantages of definition (\ref{eq:Hjf}) over other
definitions of the slow-roll parameters \cite{SL,MS2,Lea}. First of all,
the physical interpretation of the functions given by Eq.~(\ref{eq:Hjf})
and equations of motion (\ref{eq:eom}) is straightforward and model independent
(not restricted to models with a single scalar field). Secondly, the notation 
is concise, leading to significant simplification in the involved expressions. 
Thirdly, the definition is easy to memorise. The link between the first
three horizon flow functions and various definitions of the corresponding
slow-roll parameters is presented in table \ref{tab1}.
\begin{table}[t]
\begin{minipage}{\textwidth}
\begin{tabular}{ccccc}
\hline 
this & Lidsey et al. \cite{Lea} & Martin \& Schwarz \cite{MS2} & 
   Stewart \& Lyth \cite{SL}\footnote{$\epsilon \to \epsilon_1$ 
                                       and $\delta \to \delta_1$} \\
work & & & Stewart \& Gong \cite{GS} \\ 
\hline
$\epsilon_1$ & $\epsilon$ & $\epsilon$ & $\epsilon_1$ \\
$\epsilon_2$ & $2\epsilon - 2\eta$ & $2\epsilon - 2\delta$ & 
   $2\epsilon_1 + 2\delta_1$ \\
$\epsilon_2\epsilon_3$ & $4\epsilon^2 - 6\epsilon\eta + 2\xi^2$ & 
   $2\xi$ & $4\epsilon_1^2 + 6\epsilon_1\delta_1 - 2\delta_1^2 +
   2\delta_2$ \\
\hline
\end{tabular}
\caption{Conversion table for different definitions of the expansion 
basis.\label{tab1}}
\end{minipage}
\end{table}

To proceed with the calculation of the inflationary perturbations, we have
to relate the comoving Hubble rate, $aH$, to conformal time. From the
definition ${\rm d} \tau \equiv {\rm d} t/a$ we find after a partial 
integration 
\begin{equation}
\label{conf}
\tau = - {1\over aH (1 - \epsilon_1)} + \int {\epsilon_1 \epsilon_2 \over 
(1 - \epsilon_1)^2} {{\rm d} N\over a H }. 
\end{equation}

The equation of motion of the perturbations is (see \cite{MS,MS2} for the 
notation):
\begin{equation}
\label{eqmu}
\mu (k,\tau)''+ (k^2 - \frac{z''}{z}) \mu (k,\tau)=0,
\end{equation}
where a prime denotes a derivative with respect to conformal time and
where $z=a\sqrt{\epsilon_1}$ for scalar perturbations and $z=a$ for
tensor perturbations. This equation should be solved 
with the following initial conditions
\begin{equation}
\label{ini}
\lim _{k/(aH)\rightarrow +\infty}\mu _{\rm S,T}(\tau)=\mp
4\sqrt{\pi }l_{\rm Pl}\frac{e^{-ik(\tau -\tau_{\rm i})}}{\sqrt{2k}},
\end{equation}
where $l_{\rm Pl}$ denotes the Planck length (the two signs stand for scalar 
and tensor perturbations respectively). Then the power spectra can be
calculated and read
\begin{equation}
\label{pspec}
k^3P_{\zeta }=\frac{k^3}{8\pi ^2}\biggl\vert \frac{\mu _{\rm S}}{z_{\rm S}}
\biggr \vert ^2,\quad
k^3P_{h}=\frac{2k^3}{\pi ^2}\biggl\vert \frac{\mu _{\rm T}}{z_{\rm T}}
\biggr \vert^2,
\end{equation}
where $\zeta$ and $h$ stand for scalar and tensorial modes respectively.

\section{Scalar perturbations}

The potential in the scalar mode equation reads 
\begin{equation}
\frac{z''}z = a^2 H^2 \left(2 - \epsilon_1 + \frac 32 \epsilon_2 + 
\frac 14 \epsilon_2^2 - \frac 12 \epsilon_1 \epsilon_2 + 
\frac 12 \epsilon_2 \epsilon_3 \right). 
\label{sp}
\end{equation}
We can solve the mode equations by Bessel functions 
\begin{equation}
\mu = (k\tau)^{1/2}[B_1 J_\nu(k\tau) + B_2 J_{-\nu}(k\tau)] \, ,
\end{equation}
with the constants $B_1$ and $B_2$, if 
\begin{equation}
\tau^2 {z''\over z} \equiv \nu^2 - \frac 14
\end{equation}
may be approximated to be constant. This condition is met if we can 
consistently neglect the time 
derivative of $\epsilon_1$ and $\epsilon_2$, which means that we have to set
$\epsilon_1 \epsilon_2 \approx 0$ and $\epsilon_2 \epsilon_3 \approx 0$ from
equations (\ref{eq:eom}). Note that this does not imply $\epsilon_n \approx 0$.
We see from equation (\ref{conf}) that we actually need to require that 
$\epsilon_1 \epsilon_2 \Delta N < A/100\%$, where $A$ is the accuracy that 
we want to achieve and $\Delta N$ is the number of e-folds during which 
the time derivatives have to be neglected.
Dropping the above mentioned terms in Eqs.~(\ref{sp}) and (\ref{conf}) we 
find the index of the Bessel functions, which is given by
\begin{equation}
\nu \approx \frac 12 + {1\over 1-\epsilon_1} + \frac 12 \epsilon_2.
\end{equation}
In the above expression, $1/(1-\epsilon_1)$ should be 
expanded for small $\epsilon_1$ and truncated at order $n$, such that 
$\epsilon_1^n > |\epsilon_1 \epsilon_2|$. 

We can now fix $B_1,B_2$ by comparing to Eq.~(\ref{ini}) in the limit $\tau
\to \tau_{\rm i}$.
This gives
\begin{equation}
B_1 = 2\pi l_{\rm Pl} {\exp[i(\nu \pi/2 +\pi/4 + k\tau_{\rm i})]
\over \sqrt{k}\sin(\pi\nu)}, \quad B_2 = - \exp(-i\pi\nu) B_1. 
\end{equation} 
The next step is to calculate the superhorizon limit of $\mu$, which becomes
\begin{equation}
|\mu| \to {l_{\rm Pl} \pi 2^{\nu + 1} (-k\tau)^{1/2 -\nu} \over 
\sqrt{k} \sin(\pi\nu) \Gamma(1-\nu)} = 
{l_{\rm Pl} 2^{\nu + 1} (-k\tau)^{1/2 -\nu} \Gamma(\nu) \over 
\sqrt{k}}.
\end{equation}
Inserting in the definition of the power spectrum (\ref{pspec}) and eliminating
$\tau$ with help of Eq.~(\ref{conf}) we find 
\begin{equation}
\label{res1}
k^3P_{\zeta } \approx {l_{\rm Pl}^2 H^2 \over \pi \epsilon_1} 
\left[\frac 1{\pi} (2 - 2\epsilon_1)^{(2\nu -1)} \Gamma^2(\nu)\right] 
\left( k \over aH \right)^{(3 - 2\nu)}.
\end{equation}
At this point we have to fix the time $t_*$ at which we evaluate $\epsilon_m$.
We introduce the abbreviation $k_* \equiv (aH)(t_*)$. We have shown above,
that terms like $\epsilon_1 \epsilon_2$ should be neglected consistently.
However, the above result certainly still involves terms like this. In the
last step we therefore have to expand Eq.~(\ref{res1}) for $\epsilon_1$ and
$\epsilon_2$ and drop all terms that are of order $\epsilon_1 \epsilon_2$ or 
smaller. 
The crucial difference to the slow-roll approximation as used by Stewart 
and Lyth \cite{SL} is that they assume $\epsilon = \epsilon_1$ and 
$\delta = \epsilon_2/2 - \epsilon_1$ to be small simultaneously, thus they 
cannot keep higher power in their slow-roll parameters.
In our approach we can, e.g., keep terms of order $\epsilon_m^5$ if they 
are larger than $|\epsilon_1\epsilon_2|$ and $|\epsilon_2\epsilon_3|$. 
Consequently,
depending on the relative magnitude of the functions $\epsilon_1$,
$\epsilon_2$ and $\epsilon_3$ we obtain different results.

\subsection{Constant-horizon approximation}

In some inflationary models as for inflation at a maximum, the time derivative
of the Hubble distance is tiny. For this kind of models, during a certain 
number of e-folds, $\epsilon_1 \ll 1$. However, it does not necessarily mean 
that all other $\epsilon_m$ have to be small as well.
We thus define the constant-horizon approximation at order $n$ for the 
situation $|\epsilon_2^n| > \max(|\epsilon_1\epsilon_2|, 
|\epsilon_2\epsilon_3|)$, 
which means that we are allowed to include the following monomials 
in the primordial spectra: $1, \epsilon_1, \epsilon_2, \dots, \epsilon_2^n$.

Expanding Eq.~(\ref{res1}) we find at third order ($n=3$)
\begin{eqnarray}
k^3P_{\zeta } &\approx& {l_{\rm Pl}^2 H^2\over \pi \epsilon_1} \left[
a_0 + a_1 \ln\left({k\over k_*}\right) + a_2 \ln^2\left({k\over k_*}\right)
+ a_3 \ln^3\left({k\over k_*}\right) + \dots \right] \\
a_0 &=& 1 - 2(C+1) \epsilon_1 - C \epsilon_2 
        + \frac 18 (4 C^2 + \pi^2 - 8) \epsilon_2^2 \nonumber \\
    & & - \frac 1{24} \left[4 C^3 - 3C(8 - \pi^2) + 14 \zeta(3) - 16 \right] 
        \epsilon_2^3 \, ,\\
a_1 &=& - 2 \epsilon_1 - \epsilon_2 + C \epsilon_2^2 
        - \frac 18 (4 C^2 + \pi^2 - 8) \epsilon_2^3 \, ,\\
a_2 &=& \frac 12 \epsilon_2^2 - \frac 12 C \epsilon_2^3 \, ,\\
a_3 &=& - \frac 16 \epsilon_2^3 \, ,
\end{eqnarray}
where $C \equiv \gamma_{\rm E} + \ln 2 - 2\approx -0.7296$ and $\zeta(3) 
\approx 1.2021$. 
The quantity that is usually 
called the amplitude is obtained by setting $k = k_*$.
We finally calculate the spectral index 
\begin{equation}
n_{\rm S} - 1 \equiv {{\rm d}\ln k^3 P_{\zeta}\over{\rm d}\ln k}\vert_{k=k_*},
\end{equation}
which is most easily obtained from Eq.~(\ref{res1}). It reads
$n_{\rm S} - 1 = - 2 \epsilon_1 - \epsilon_2$ at any order $n$ in the constant
horizon approximation. In the corresponding limits this result agrees with 
the usual slow-roll expression. Consistently with our assumptions, there is 
no ``running'' of the spectral index. 

\subsection{Growing-horizon approximation}

Power-law inflation ($a \propto t^p$) is one of the few models of inflation 
for which an exact expression for the primordial spectrum can be obtained in 
closed form. Our horizon flow functions in this case are 
$\epsilon_1 = 1/p = const.$ 
and $\epsilon_m = 0$ for $m > 1$. This observation suggests that inflationary
models with $|\epsilon_m| < \epsilon_1$ for $m > 1$ might be approximated 
by
keeping all terms in $\epsilon_1$ up to order $n$, where $n$ is the maximal
integer for which $\epsilon_1^n > 
\max(|\epsilon_1 \epsilon_2|,|\epsilon_2 \epsilon_3|)$ holds true. 
For this kind of models,
$\epsilon_1 \simeq const.$, implies that the Hubble radius grows almost
like a linear function of time.  
Hence, we define the (linearly) growing-horizon approximation to order $n$ to 
include the following terms: $1,\epsilon_1, \dots, \epsilon_1^n, \epsilon_2$.

The scalar power spectrum at third order in the growing horizon 
approximation reads:
\begin{eqnarray}
k^3P_{\zeta } &\approx& {l_{\rm Pl}^2 H^2\over \pi \epsilon_1} \left[ 
b_0 + b_1 \ln\left({k\over k_*}\right) + b_2 \ln^2\left({k\over k_*}\right) 
+ b_3 \ln^3 \left({k\over k_*}\right) + \dots \right] \\
b_0 &=& 1 - 2(C+1) \epsilon_1 
        + \frac 12 \left[4C(C+1) + \pi^2 - 10\right] \epsilon_1^2  \nonumber \\
    & & -\ \frac 13 \left[4C^3 + 3C(\pi^2 -12) + 14\zeta(3) - 19\right] 
          \epsilon_1^3 - C \epsilon_2 \, ,\\ 
b_1 &=& - 2 \epsilon_1 + 2(2C+1)\epsilon_1^2 
        - (4 C^2 + \pi^2 - 12) \epsilon_1^3 - \epsilon_2 \, ,\\
b_2 &=& 2 \epsilon_1^2 - 4 C \epsilon_1^3 \, ,\\
b_3 &=& - \frac 43 \epsilon_1^3 . 
\end{eqnarray}
Ignoring the monomials 
$\epsilon_1^2$ and $\epsilon_1^3$, but keeping $\epsilon_2$, we recover the 
result of Stewart and Lyth \cite{SL}. The spectral index at $n$-th order reads
\begin{equation}
\label{eq:GHn_S}
n_{\rm S} - 1 = 3 - 2 \nu \approx 
- 2 (\epsilon_1 + \epsilon_1^2 + \epsilon_1^3 + \dots + \epsilon_1^n) 
- \epsilon_2 ,
\end{equation}
and again there is no ``running'' of the spectral index. 
   
\section{Tensor perturbations}

Now the potential in the mode equation reads
\begin{equation}
\label{potT}
\frac{z''}z = (aH)^2 \left(2 - \epsilon_1 \right),
\end{equation}
which allows us to solve the mode equation with the same approximation as in 
the scalar case, but with Bessel function index
\begin{equation}
\nu \approx \frac 12 + {1\over 1 - \epsilon_1} \ .
\end{equation}
The following steps are analogous to those for the calculation of the scalar 
perturbations, the basic difference is that no contribution of $\epsilon_2$ 
arises in any case.

Using the constant-horizon approximation we obtain
\begin{eqnarray}
k^3P_h &=& {16 l_{\rm Pl}^2 H^2\over \pi} \left[
\mathtt{a}_0 + \mathtt{a}_1 \ln\left({k\over k_*}\right) + \dots \right] ,\\
\mathtt{a}_0 &=& 1 - 2(C+1) \epsilon_1 ,\\
\mathtt{a}_1 &=& - 2 \epsilon_1 , \\
n_{\rm T} &=& -2 \epsilon_1 , 
\end{eqnarray}
at any order. Note that all $\mathtt{a}_i$ for $i \geq 2$ vanish in the
constant horizon approximation and that there are no higher order corrections 
to $n_{\rm T}$, which follows from the absence of any $\epsilon_2$-dependence 
in the potential (\ref{potT}) and the fact that higher powers of $\epsilon_1$
are consistently neglected. The growing-horizon approximation leads to
\begin{eqnarray}
k^3P_h &=& {16 l_{\rm Pl}^2 H^2\over \pi} \left[
\mathtt{b}_0 + \mathtt{b}_1 \ln\left({k\over k_*}\right) 
+ \mathtt{b}_2 \ln^2\left({k\over k_*}\right)
+ \mathtt{b}_3 \ln^3 \left({k\over k_*}\right) \dots \right] \\
\mathtt{b}_0 &=& 1 - 2(C+1) \epsilon_1
        + \frac 12 \left[4C(C+1) + \pi^2 - 10\right] \epsilon_1^2  \nonumber \\
    & & -\ \frac 13 \left[4C^3 + 3C(\pi^2 -12) + 14\zeta(3) - 19\right]
        \epsilon_1^3 ,\\
\mathtt{b}_1 &=& - 2 \epsilon_1 + 2(2C+1)\epsilon_1^2
        - (4 C^2 + \pi^2 - 12) \epsilon_1^3 ,\\
\mathtt{b}_2 &=& 2 \epsilon_1^2 - 4 C \epsilon_1^3 ,\\
\mathtt{b}_3 &=& - \frac 43 \epsilon_1^3 , 
\end{eqnarray}
at third order, and 
\begin{equation}
\label{eq:GHn_T}
n_{\rm T} = -2 (\epsilon_1 + \epsilon_1^2 + \epsilon_1^3 
+ \cdots + \epsilon_1^n),
\end{equation}
at $n$th order. 
Again both results are consistent with the results of Stewart 
and Lyth \cite{SL}.

\section{Consistency relations}

It is interesting to inspect the so-called `consistency relations of 
inflation' for models that do not belong to the class of slow-roll inflation.
We define the tensor-to-scalar ratio $r \equiv P_h/P_\zeta$. The 
`classic'
result for slow-roll models reads at first order $n_{\rm T} = - r/8$. We find 
for the constant horizon and for the growing horizon approximations $r = 16
\epsilon_1$ at any order. This result may be used to express the corresponding 
expressions for the tensorial spectral index as 
\begin{equation}
n_{\rm T} = - \frac{r}{8} 
\end{equation} 
for the constant horizon approximation at any order, and 
\begin{equation}
\label{eq:ConsRel}
n_{\rm T} = - 2 \left[ \left(\frac{r}{16} \right) + 
\left(\frac{r}{16} \right)^2 
+ \dots + \left(\frac{r}{16} \right)^n \right]
\end{equation}
for the growing horizon approximation at order $n$. Both results are consistent
 with the second order result of Ref. \cite{Lea}, $n_{\rm T}  = -2[ (r/16) -
(r/16)^2 - (n_{\rm S} - 1) (r/16)]$, since $(n_{\rm S} - 1) (r/16) = 
- 2 (r/16)^2 + {\cal O}(\epsilon_1 \epsilon_2)$. For models where the 
constant horizon approximation applies we expect that $r \ll 1$, thus the 
prospects to detect the tensors and therefore to test the consistency 
relation are bad. In the case of those models for which the growing horizon 
approximation is suited, there is a chance to detect the tensor contribution.
Here, the corrections to the slow-roll results given by Eq.~(\ref{eq:ConsRel})
may be relevant. However, current data seem to indicate that $r < 1$, which 
implies that higher order corrections to the classic consistency relation are 
not important.

\section{Conclusion}
\begin{figure}
\setlength{\unitlength}{\linewidth}
\begin{picture}(1,0.8)
\put(0.54,0.04){\makebox(0,0){$\epsilon_1$}}
\put(0.1,0.4){\makebox(0,0){$\vert\epsilon_2\vert$}}
\put(0.5,0.4){\makebox(0,0){\epsfig{figure=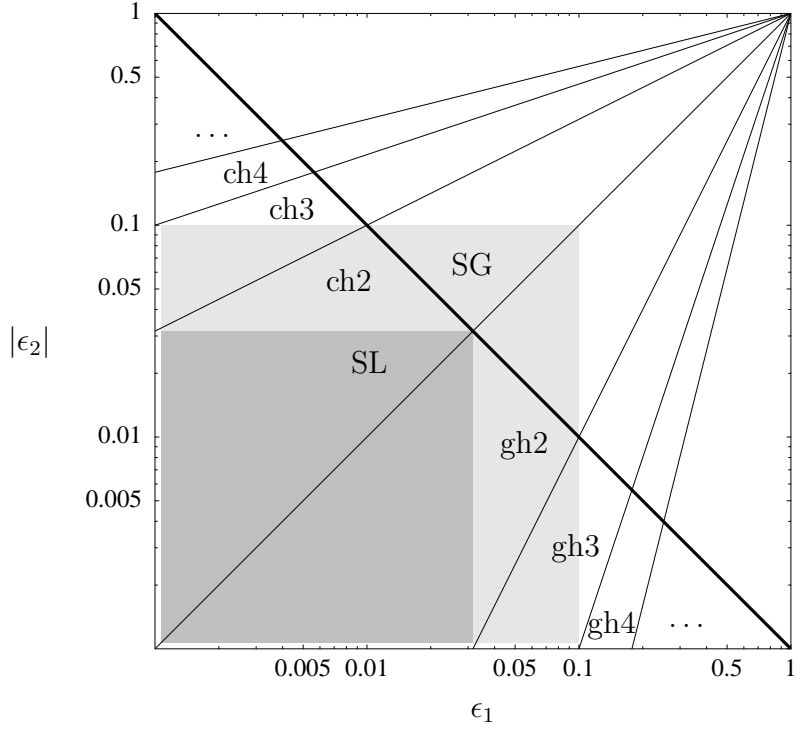,width=0.7\linewidth}}}
\put(0.43,0.38){\makebox(0,0){SL}}
\put(0.53,0.475){\makebox(0,0){SG}}
\put(0.41,0.46){\makebox(0,0){ch2}}
\put(0.355,0.53){\makebox(0,0){ch3}}
\put(0.31,0.565){\makebox(0,0){ch4}}
\put(0.28,0.6){\makebox(0,0){\dots}}
\put(0.58,0.3){\makebox(0,0){gh2}}
\put(0.63,0.2){\makebox(0,0){gh3}}
\put(0.665,0.125){\makebox(0,0){gh4}}
\put(0.74,0.125){\makebox(0,0){\dots}}
\end{picture}
\caption{ Regions in the $\epsilon_1$-$|\epsilon_2|$ parameter space where the 
spectral amplitudes can be calculated with an accuracy better than $1\%$.
In the dark shaded region the Stewart-Lyth (SL) approximation \cite{SL}, as 
well as all other approximations are fine. Second-order corrections,
as calculated by Stewart and Gong (SG) \cite{GS}, extend that region to the 
light shaded region. The constant horizon approximation at order $n$ (ch$n$), 
and the growing horizon approximation at order $n$ (gh$n$), do well below
the thick line. The rays indicate where the corresponding higher order 
corrections are necessary.
The thick line itself is the condition $\epsilon_1 |\epsilon_2| < 
(A/100\%)/\Delta N$, with $\Delta N =10$ and $A = 1\%$. As is easily seen the 
ch2 and gh2 regions are included within the SG region.
The ch$n$ and gh$n$ regions with $n > 2$ allow us to go beyond the SG 
approximation.}
\end{figure}

We may compare our result with the work by Stewart and Gong \cite{GS}.
They used a new method to obtain the scalar amplitude at second 
order in the slow-roll parameters. In our notation they obtain:
\begin{eqnarray}
k^3P_{\zeta }|_{k=k_*} 
&\approx& {l_{\rm Pl}^2 H^2\over \pi \epsilon_1} \Biggl\{
1 - 2(C+1) \epsilon_1 - C \epsilon_2 \nonumber \\
& & + \left[2C(C+1) + \frac{\pi^2}{2} - 5\right] \epsilon_1^2 
+ \left(\frac 12 C^2 + \frac{\pi^2}{8} - 1\right) \epsilon_2^2 \nonumber \\
& & + \left[C(C-1) + \frac{7 \pi^2}{12} - 7\right] \epsilon_1 \epsilon_2
+ \left(- \frac 12 C^2 + \frac{\pi^2}{24} \right) \epsilon_2 \epsilon_3
\Biggr\}, 
\end{eqnarray}
which agrees with our result if $\epsilon_1 \epsilon_2 \approx 0$ and 
$\epsilon_2\epsilon_3 \approx 0$. 

Our results can be applied to many interesting models of inflation.  
The constant horizon approximation works fine if the inflaton field sits 
close to a maximum of the potential. On the other hand the growing horizon 
approximation is a good approximation for chaotic inflation models with 
monomial potential $\propto \phi^p$, when $p > 4$. To give an example, for
$p = 8$ we have $\epsilon_1 \approx 0.04$ and $\epsilon_2 \approx 0.02$, 
which allows us to take all quadratic terms in $\epsilon_1$ into account.  
We plot in figure 1 the domain of applicability of the various approximations
in the $\epsilon_1$-$|\epsilon_2|$ plane. We tacitly assume that $|\epsilon_3| 
< \min[\epsilon_1,|\epsilon_2|]$, for the sake of simplicity of the argument. 
If this condition is not satisfied an analogous three dimensional plot has to 
replace figure 1.

We are grateful to A.~R.~Liddle, S.~Leach and J.~Martin for useful 
discussions. 
D.J.S.~thanks the Austrian Academy of Sciences for financial support.  
The work of C.A.T.-E.~and A.A.G.~is supported in part by the CONACyT 
grant 32138--E and the Sistema Nacional de Investigadores (SNI).

\end{document}